\documentclass{article}
\usepackage{graphicx}
\usepackage{amssymb}

\begin{document}

\begin{center}
\Large
{\bf Dzyaloshinski-Moriya interactions in the kagom\'e lattice}

\vspace{5mm}
\normalsize
Maged~Elhajal\footnote{E-mail: elhajal@polycnrs-gre.fr}, Benjamin~Canals, Claudine~Lacroix

\vspace{5mm}
\small
Laboratoire Louis N\'eel, 25 avenue des Martyrs, 
BP 166, 38042 Grenoble Cedex 9, France

\vspace{5mm}
August 2001

\vspace{5mm}
\normalsize

\end{center}

\begin{abstract}
The kagom\'e lattice exhibits peculiar magnetic properties due
to its strongly frustated cristallographic structure, based on corner sharing
triangles.
For nearest neighbour antiferromagnetic Heisenberg interactions
there is no N\'eel ordering at zero temperature both for quantum and classical spins.
We show that, due to the peculiar structure, antisymmetric Dzyaloshinsky-Moriya
interactions (${\bf D} . ({\bf S}_i \times {\bf S}_j)$) are present in this lattice.
In order to derive microscopically this interaction we consider a set of 
localized $d$-electronic
states.
For classical spins systems, we then study the phase diagram $(T, D/J)$ through
mean field approximation and 
Monte-Carlo simulations and show that the antisymmetric
interaction drives this system to ordered states as soon as this interaction is
non zero.
This mechanism could be involved to explain the magnetic structure of 
Fe-jarosites.

\end{abstract}

\section{Introduction}

\begin{figure}[h]
\centering \includegraphics[height=3cm]{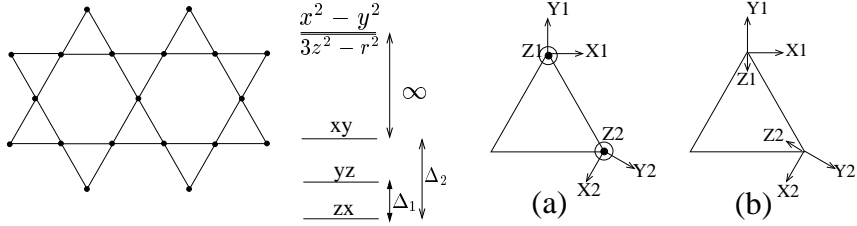}
\caption{kagom\'{e} lattice (left). Cristal  field scheme (x, y and z refer 
to the local axes) and the two different cases considered for the derivation 
of the DMI.}
\label{cef}
\end{figure}

The kagom\'{e} lattice (Fig. \ref{cef}) with nearest neighbour
antiferromagnetic exchange interactions exhibits peculiar magnetic
properties, due to geometrical frustration \cite{Revue}.
For classical spins it remains disordered
within mean field approximation, reflecting the macroscopic degeneracy of
the ground state.
Thermal or quantum fluctuations may leave
partially this degeneracy, leading to `order by disorder'.
Here we show
that Dzyaloshinski-Moriya interactions (DMI) are present in this system due
to the low local symmetry; we discuss the effect of DMI in relation with the
magnetic structure observed in Fe-jarosite \cite{Andrew-Fe}. 

The DMI between two spins \textbf{S}$_{i}$ and \textbf{S}$_{j}$ was first
introduced phenomenologically, using symmetry considerations, by
Dzyaloshinski \cite{Dzyaloshinski}, and Moriya showed that it arises from
taking into account the spin-orbit coupling ($\lambda \mathbf{L}.\mathbf{S}$%
) in the super-exchange interaction between localized magnetic electrons %
\cite{Moriya}. Including DMI, the hamiltonian is written as:

\begin{equation}
\label{hamiltonian}
H=J\sum_{\langle i,j\rangle}{\bf S}_i.{\bf S}_j+\sum_{\langle i,j\rangle}
{\bf D_{ij}}.\left({\bf S}_i\times{\bf S}_j\right)\hspace{5mm} (J>0)
\end{equation}

The allowed \textbf{D} vectors of the DMI are restricted by the symmetry of
the cristal (Moriya's rules \cite{Moriya}). 
In the kagom\'e lattice, there is
no inversion symmetry between 2 neighboring sites (see figure \ref{cef}) and
the symmetry of the simple kagom\'{e} lattice allows for \textbf{D}
perpendicular to the kagom\'{e} plane. 
However, in real compounds such as
jarosites (see Ref. \cite{Andrew-Fe}), the non-magnetic atoms (which are not on the 
kagom\'{e} lattice) are also relevant, as they contribute to the cristal
field and to the super-exchange interaction.
Consequently they may lower the symmetry
and allow for other directions of \textbf{D}. 

In any case, applying Moriya's rules will only \emph{restrict} the possible 
\textbf{D}, but is not a proof of their existence. This is the reason for
the microscopic derivation of the DMI in section \ref{calcul_microscopique}.

\section{Microscopic derivation of Dzyaloshinski-Moriya interaction}
\label{calcul_microscopique} 
We have derived the DMI starting from an
(arbitrary) set of localized d electronic states with the cristal
electric field shown in Fig. \ref{cef}a with one electron per magnetic
site. \\

Following Moriya's formalism \cite{Moriya} we find:

\begin{equation}
D_x=D_y=0\hspace{5mm}D_z=-\frac{\lambda\sqrt3}{U\Delta_1}\left( 
\frac{\left(dd\pi\right)^2}{2}-\left(dd\pi\right)\left(dd\delta\right)-\frac{3}{2}\left(dd\delta\right)^2\right) 
\label{vecteur_D}
\end{equation}

where U is the on-site Coulomb repulsion, and $dd\delta $ and $dd\pi $ are the
hopping integrals as defined in \cite{Slater}. 

In the jarosites (Ref. \cite{Andrew-Fe}), each magnetic site is surrounded by an
octahedron of oxygen atoms which gives rise to the cristal electric field.
The z axis of these octahedra is tilted towards the centers of the triangles
of the kagom\'{e} structure by an angle $\alpha $. 
In order to describe
these compounds, we have derived the DMI as above but with the local axis on
each magnetic site represented on Fig. \ref{cef}b. 
In this case, the \textbf{D}$_{ij}$ are in the plane perpendicular to the bond i-j. 
The
expressions of $D_{x},$ $D_{y},$ $D_{z}$ can be obtained as in (\ref{vecteur_D}); 
they depend on $\alpha $ but we do not give them here for
simplicity. 
In the following, we call $\beta $ the angle between \textbf{D}$_{ij}$ and the
z axis.

\section{Low temperature magnetic structure}
Looking at the ground state within mean field theory, we observe in both
cases ($\beta =0$ and $\beta \neq 0$) a long range ordered structure.
It is a \textbf{q}=\textbf{0} structure (the magnetic cell
contains 3 sites). 
In the case of \textbf{D} perpendicular to the kagom\'{e}
plane ($\beta=0$ or $\pi$), all spins lie in the kagom\'{e} plane, and only 
one degree of freedom
remains, which corresponds to a `global' rotation of all spins in the plane.
In this case the DMI not only acts as an effective easy-plane anisotropy, it
also selects one chirality for all the triangles, depending on the sign of D$_{z}$. 

\begin{figure}
\centering \includegraphics[height=5cm]{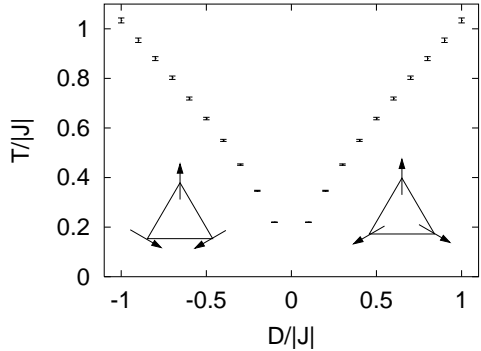}
\caption{Phase diagram for {\bf D} perpendicular to the kagom\'{e} plane.}
\label{phase_diag}
\end{figure}

In the case of {\bf D} not perpendicular to the kagom\'{e} plane 
($\beta\neq 0$), the effect of the in-plane component of {\bf D} is the 
following :
the spins are no longer coplanar, each spin is directed in one fixed
direction whose projection on the kagom\'{e} plane is towards the centers of
the triangles (the chirality on the right of Fig. \ref{phase_diag} is selected) and 
all the spins have the \emph{same} out of plane
component giving rise to a weak ferromagnetic magnetization perpendicular to
the kagom\'e plane.  
This magnetization increases with the in-plane component
of \textbf{D} and decreases when $\frac{J}{D}$ increases as it should
because isotropic exchange (J) selects coplanar magnetic structures. 
If D$_z>0$, it selects the same chirality as the in-plane component of 
{\bf D}, and its effect is to reduce the out of plane magnetization. 
D$_z<0$ tends to select the other chirality (on the
left of Fig. \ref{phase_diag}), resulting in a competition with the 
in-plane component of {\bf D}, and there is a critical value of the
angle $\beta $ between {\bf D} and the z axis, above which the structure
becomes coplanar. 

In fact such a non-coplanar magnetic structure has been observed in
Fe-jarosite (Ref. \cite{Andrew-Fe}) and could be due to DMI. 
In this jarosite it is also
observed that the structure becomes coplanar at lower temperature which
could be due to a small structural change, or 
possibly due to interplane interactions which
are not taken into account in this work. 

In order to study the effect of temperature on this magnetic structure, we
have performed Mont\'{e}-Carlo simulations with classical Heisenberg spins
for different vectors \textbf{D}. 
In all cases we observe a phase transition.
The critical temperature has been obtained as a function of $\frac{D}{J}$, 
which is represented on Fig. \ref{phase_diag} for the case D
perpendicular to the plane.
It is seen that the critical temperature is of the order of D. 

Thermal fluctuations also lift the degeneracy of the system without DMI 
\cite{Chalker}\cite{Reimers}, but there is no caracteristic energy scale, 
i.e, this selection occurs asymptotically when $T \rightarrow 0$.
On the other hand, the phase transition due to DMI occurs at a  
temperature of the order of D.
Therefore, we expect that DMI when present will always dominate the low 
temperature behaviour.

In conclusion, we have shown that a DMI is relevant for the kagom\'{e}
lattice and found that it can induce a phase transition at a temperature of
the order of D. It could explain some magnetic structures observed in the
Fe-jarosites \cite{Andrew-Fe}.

\end{document}